# Theory of tectonics in the sphere

António Ribeiro
Laboratório de Tectonofísica e Tectónica Experimental, Bloco C2, 5º piso, Campo Grande, 1700 Lisboa, Portugal

Luis Matias
Centro de Geofísica da Universidade de Lisboa, Campo Grande, Ed. C8, piso 6, 1749-016 Lisboa, Portugal

Rui Taborda
Laboratório de Tectonofísica e Tectónica Experimental, Bloco C2, 5º piso, Campo Grande, 1700 Lisboa, Portugal

**Abstract**

Soft or Deformable Plate Tectonics in the sphere must follow geometric rules inferred from the orthographic projection. An analytic equivalent of this geometry can be derived by the application of Potential Field Methods in the case of Atlantic type oceans. Laplace equation must be obeyed by the velocity field between the ridge and the passive margin if we neglect the very slight compressibility of ocean lithosphere. A strain wave propagates in the sphere analogous to the behaviour of a free harmonic oscillator. Combining zonal harmonics of order one and sectorial harmonics of degree one we obtain a tesseral harmonic equivalent to the orthographic solution. This potential field approach is valid for homogeneous deformation regime in oceanic lithosphere. Above a compression threshold of 5 to 10% buckling and simultaneous faulting occurs. In Pacific type oceans a dynamic approach, similar to a forced oscillation, must be applied because there are sinks in subduction zones.

## 1. Introduction

A theory of deformable or soft plate tectonics was recently proposed by Ribeiro (2002, pp. 60-86). It was demonstrated that deformable plate rotation in the sphere must follow geometric rules inferred from orthographic projection. The purpose of this text is to show that an analytic equivalent of this geometry can be derived by the application of Potential Field Methods. First, we will summarise the basis for the geometric model of plate deformation (Ribeiro, 1993; Ribeiro, 2002, pp. 60-86).

## 2. Geometric Approach to Tectonics in the Sphere

Plates move by rigid rotations around Euler Poles according to standard Plate Tectonics Theory. The symmetry of rigid plate rotation is cylindrical or axis-symmetric, lower than the spherical symmetry of the globe. This symmetry can be represented by a cylinder with a circular section representing the rigid rotation, such as the one produced by the oblique Mercator projection with an axis through the relative Euler Pole of the plates involved.

If plates suffer intraplate deformation during their rotation, there is a symmetry breaking. With intraplate deformation, the cylinder representing the plate movement looses symmetry because the circular section becomes elliptic. Assuming a constant Earth Radius, R, because the altitude or depth variations are negligible when compared to plate dimensions, we postulate that soft plate rotations must obey the following rules:

Rule 1: Deformation must be zero at the mid-ocean ridge.

Rule 2: Transform faults must remain small circles in the sphere or parallel straight lines (in the oblique Mercator projection) in oceanic domains of the same age.

Rule 3: The lithosphere must suffer plane deformation.

These rules guarantee that the plate does not suffer torsion along horizontal and vertical axes.

Orthographic projection (De Paor, 1983) allows us to represent plate deformation that obeys the previous rules. In fact, the orthonet is the only projection that can be used in tensor manipulation, such as the estimation of strain, because the great circles of the orthographic projection are ellipses with an axial ratio that varies from 1, a circle, to ∞, a line, as a function of their orientation relative to the plane of projection. An oblique transverse orthographic projection represents plate deformation with a pole around the Euler Pole of rotation and the mid-ocean ridge (MOR) projected as the central reference meridian. In this case, small circles project as straight lines, representing the transform faults that guide the plate rotation. Considering that $\phi_0$ represents the angular distance from the MOR on the undeformed plate (see figure 1) and $\phi$ the same distance on the deformed plate, then from orthographic projection rules we have

$$\phi = \sin \phi_0 \qquad (1)$$

The axial ratio, $R=(1+e_1)/(1+e_2)$, of a strain ellipse at the surface of the Earth will be only a function of the angular distance from the MOR, $\phi_0$. At the equator of the projection $e_1=0$ and

$$1+e_2 = \frac{d\phi}{d\phi_0} = \cos\phi_0 \quad <1$$

We obtain then that the strain ratio at the equator is the inverse of the cosine of angular distance to the MOR on the undeformed plate,

$$R = \frac{1+e_1}{1+e_2} = (\cos\phi_0)^{-1} \qquad (2)$$

These same results can be inferred from a heuristic approach. In a steady-state spreading ocean, the two plates diverge symmetrically around a fixed Euler pole. In the undeformed state, the arc described by the symmetrical spreading ocean, $\phi_0$, is a function of the half spreading rate at the mid-ocean ridge, $\dot{\phi}_0$, and time $t$, $\phi_0 = \dot{\phi}_0 t$.

In the deformed state the stretch is variable, as a function of time

$$\frac{\phi}{\phi_0} = F(t)$$

In the ridge there is no deformation, hence F(0)=1.

So we must impose that

$$1 \equiv \lim_{t \to 0} F(t) = \lim_{\phi_0 \to 0} \frac{\phi}{\phi_0} = \lim_{\phi_0 \to 0} \frac{\sin \phi_0}{\phi_0}$$

Therefore the solution $\phi = \sin \phi_0$ (equation 1 from the orthographic projection) satisfies the required boundary conditions at the ridge. Considering the influence of the time, we have also

$$\phi = \sin(\dot{\phi}_0 t) \qquad (3)$$

All the previous equations are Lagrangian descriptions of the deformation suffered by the plate at the projection equator.

For a line parallel to plate motion at $\theta$ degrees from the Euler pole we will use a coordinate frame where the Euler Pole is at North or South pole of the projection and the ridge is the central reference meridian. A point in the sphere is referred by its angular distance to the Euler Pole, the colatitude θ, and ϕ or ϕ₀, the angular distance to the ridge crest (see figure 2).

The Plate Kinematics theory requires that the linear displacement $D_\theta$ at the colatitude θ, is given by $D_\theta = D \sin \theta$, where D is the linear displacement at the equator. Given that on the deformed state we have $D = R\phi = R\sin(\dot{\phi}_0 t)$, where R is the earth radius, we obtain for the linear displacement at colatitude θ the expression

$$D_\theta = R \sin \theta \, \sin(\dot{\phi}_0 t) \qquad (4)$$

Equation 4 contains equation 3 as a special case when $\theta = 90°$, (Ribeiro, 2002, pp. 60-86).

The velocity field, or the instantaneous spatial variation of the displacement field, across a moving plate can be expressed by plotting the angular displacement rate as a function of angular distance to mid-ocean ridge (ibid.). With this procedure we eliminate the variation due to the angular distance to the Euler Pole, $\theta$, for baselines parallel to plate motion. The angular displacement ϕ₀ during time interval t will be, in rigid rotation, $\phi_0 = \dot{\phi}_0 t$, with $\dot{\phi}_0$ being the angular half-spreading rate, in radians per unit of time. In rigid plate this is a constant angular velocity. In deformable plate theory the velocity, or displacement rate, of a point along the base line, $\dot{\phi}$, is

$$\dot{\phi} = \frac{d\phi}{dt} = \frac{d[\sin(\dot{\phi}_0 t)]}{dt} = \dot{\phi}_0 \cos\phi_0 = \dot{\phi}_0 \sqrt{1-\phi^2} \quad (5)$$

This is a Eulerian equation because $\phi$ is measured in the deformed state and $\dot{\phi}_0 = \dot{\phi}$ at the ridge.

We can compare the real geodetic data that gives us linear or angular displacement rates with the theoretical values for rigid plate theory, estimated from kinematic models such as NUVEL 1A, with the corresponding theoretical values applying the deformable plate theory. If the real values are closer to theoretical values using deformable plate theory than to theoretical values using rigid plate theory, the deformable plate theory is validated. For the Atlantic Ocean domain, estimating a maximum half-extension of 25º, equation (5) predicts a 10% difference between the velocities measured at the MOR and those measured from points beyond the ocean margins.

Thus, the preceding theory can be used to test the soft plate model, by comparing real displacements obtained by geodetic methods and by theoretical displacements predicted by a rigid plate kinematic models, assuming quasi steady-state motion of plates in the last 3 MA (Ribeiro, 2002, pp. 60-86). We concluded that the oceanic lithosphere is not rigid and is more easily deformed than cratons. For this reason we can use geodetic methods in oceans to test the geometric rules of orthographic projection and their analytical equivalent to potential field methods in Soft Plate theory.

We will first examine the simple case of Atlantic type oceans and afterwards the more complex case of Pacific type oceans.

### 3. Atlantic type oceans: potential field approach

In Atlantic type oceans, the main driving force for Plate Tectonics is ridge push. Assuming that resistance to ridge push is proportional to the distance to the ridge, we can assimilate the ridge system to a free harmonic oscillator (Ribeiro, 2002, pp. 60-86). The differential vector field equation describing the displacement follows a cosine rule. As the ocean spreads, there is a longitudinal strain wave that propagates between the ridge and the passive margin. Beyond that, the continent behaves rigidly. The compression expressed by the propagating wave suggests that it must follow the theoretical scheme of spherical harmonics, a branch of Theoretical Physics that has been successfully applied to Earth Science.

In fact, the Laplace equation should apply to the deformation of oceanic lithosphere in an Atlantic type ocean for the following reasons:

First, between the ridge and the passive margin, mass is preserved in the oceanic lithosphere. Volume is not preserved because the slowly cooling oceanic lithosphere is slightly compressible at the million-year scale of time. Nevertheless the Laplace equation should be a good approximation to the displacement field at the scale of 1 to 100 years of observation by geodetic methods, both classical and spatial. At this scale of time, the very small displacements at cm or mm in base lines of thousands of kilometres are equivalent to very small displacement rates and the condition of incompressibility is a very good approximation. At this scale of observation the data cannot distinguish between elastic and viscous rheology for the lithosphere. For this purpose we need longer periods of observation, in order to integrate the differential displacements over time.

Second, deformation must be the same at the top and the base of the lithosphere. Therefore, the flow must be irrotational: homogenous shortening by pure compression is the response to stress distribution in the lithosphere. The plate rotates around an Euler Pole situated outside the Plate. This condition is not observed in a spinning plate, rotating around an Euler pole inside the plate.

Third, for each stage pole the plate position is uniquely determined and is only a function of time if angular velocity is constant. The instantaneous flow pattern depends only on the instantaneous boundary conditions and not on the previous history of flow (Tritton, 1988, pp. 115-116).

These arguments show that the displacement field should derive from a velocity potential function obeying Laplace's equation. The harmonic solution has a maximum at the ridge crest and a minimum at the contact of the oceanic and continental lithosphere along the passive margin of the Atlantic type ocean.

Spherical harmonics appear in several branches of Theoretical Physics, in particular when solving the Laplace equation by separation of variables in spherical coordinates (r,θ,φ), where r is the radius, θ is the co-latitude and φ the longitude, as defined before. The spherical harmonics are represented by $Y_{lm}$ (see Pipes and Harvill, 1971, pp. 338-340, for example) and their order and degree are defined by two integers where $l \geq 0$ and $-l \leq m \leq l$.

For positive values of *m*, the spherical harmonics may be defined in terms of the associated Legendre functions of the same order,

$$Y_{lm}(\theta,\phi) = A_{lm} P_l^m(\cos\theta) e^{im\phi}$$

where $A_{lm}$ is a normalization factor,

$$A_{lm} = \sqrt{\frac{2l+1}{4\pi} \frac{(l-m)!}{(l+m)!}}$$

The associated Legendre function $P_l^m(x)$, with $x = \cos\theta$, is defined in terms of the ordinary Legendre polynomials by

$$P_l^m(x) = (-1)^m (1-x^2)^{m/2} \frac{d^m}{dx^m} P_l(x)$$

Plate deformation introduces a component of longitudinal variation in vector fields that does not exist in a rigid plate rotation. But even in rigid plate rotation there is another component of latitudinal variation: linear velocity is a sinusoidal function of angular distance to the Euler pole. It has a maximum value at π/2 of the Euler pole and a minimum value of 0 at the Euler pole itself. For the moment we neglect any radial component because altitude variations are small when compared to latitude and longitude variations.

Spherical harmonics obey spherical symmetry for boundary conditions and can describe both components. A zonal component (m=0) with order $l$ describes the latitudinal variation. The sectoral component with degree $|m|=l$ describes the longitudinal variation. Both components are generally present and describe a tesseral harmonic. In the case of plate tectonics, the zonal component is of order 1, because there is only one maximum at the Euler equator and the sector component is also of order 1, because there is also one maximum at the maximum angular distance to the ridge crest.

The lower order Legendre polynomials are $P_0=1$ and $P_1=x=\cos\theta$. Therefore, $P_1(\theta)$ describes the latitudinal variation as a function of angular distance to Euler pole, $\theta$, in the form of a zonal harmonic.

In deformable plate tectonics there is also longitudinal variation as a function of angular distance to ridge crest, $\phi_0$. The spherical harmonic of order 1 and degree 1 is given by

$$Y_{11} = -\sqrt{\frac{3}{8\pi}} \sin\theta \, e^{i\phi_0}$$

Its imaginary part

$$S_{11} = K \sin\theta \sin\phi_0$$

is equivalent to the orthographic solution presented before (equation 4), for a spherical earth with radius K. This equation describes the displacement field $D_\theta$ on a spherical earth where plates move by rotations around the Pole and deform obeying the Laplace's equation.

This solution expresses clearly the symmetry of the deformed plate. Rigid plate rotation obeys axis-symmetry or cylindrical symmetry. Deformable plate rotation breaks this symmetry because the cylinder representing the symmetry of the process changes from a circular to an elliptic section, as stated before. The analytic basis to these symmetry arguments is the fact that circular harmonics admit the equation of simple harmonic motion as the solution for the functional dependence of $\phi_0$, angular distance to ridge crest (Pipes and Harvill, 1971, pp. 338-340).

The previous model is valid for instantaneous displacement field across the spreading Atlantic-type oceans, but this ocean evolves in geological time in the order of 100-200 MA, so at any instant we see the finite effect of this evolution. The orthographic model (Ribeiro, 2002, pp. 60-86) shows that the oceanic lithosphere spreads at the ridge and is compressed across the whole Atlantic-type ocean in the homogenous deformation regime. For slow spreading oceans, the strain in the passive margin is significantly different from zero but always moderate. The result is that Atlantic-type, slow spreading oceans, have full angular widths lower than π/2, showing that, as deformation progresses with time, buckling and whole-lithosphere failure lead to subduction and replace the previous homogenous deformation regime in the oldest segments of oceanic lithosphere. This change in regime is due to the fact that the compressibility of oceanic lithosphere is limited as a function of material properties. For strains above a certain threshold, in the order of 5 to 10%, compression changes from homogenous shortening to buckling and simultaneous faulting (Ribeiro, 2002, pp. 60-86).

## 4. Pacific type oceans: Dynamic Approach

In a Pacific type ocean, mass is not preserved between ridge crest and the subduction zones that encircle the ocean. Therefore the Laplace equation should not apply to the displacement field.

Ribeiro (2002, pp. 60-86) proposed that in this case the dynamics of the evolution of the lithosphere should be similar to a forced oscillation involving forces of two origins: ridge push resulting from free oscillation at the ridge crest and slab pull induced by the presence of a subduction zone. The differential linear equation governing this case is unknown at the present time, but symmetry arguments suggest that the range of slab pull must be $\pi$ instead of the range of ridge push which is complete at $\pi/2$. This phase difference should induce the forced oscillation. In fact, we can imagine an earth with a ridge and at $\pi$ angular distance a subduction zone around a unique Euler Pole. Slab pull must act during the whole range of oceanic lithosphere that covers this oceanic earth entirely. In fact, the Pacific Plate has an angular width higher than $\pi/2$, in accordance with the previous views.

In North Atlantic and Pacific type oceans, the sea floor depth is controlled by the slow cooling of the oceanic lithosphere (Parsons and Sclater, 1977). This means that there is a scalar field corresponding to temperature variation with depth governed by the heat conduction equation. This scalar field can explain sea floor depth where oceanic lithosphere is in isostatic equilibrium, and not reheated by plume events. The depth/age correlation is slightly different between the N. Atlantic and the N. Pacific oceans. Slightly deeper ($\approx 0.5$ km) sea floor depths occur in the N. Atlantic when compared with ocean bottoms of the same age in the N. Pacific (Turcotte and Schubert, 1982, pp. 181-182). This may be interpreted as representing a slight thinning of the Pacific Plate produced by the additional tensional force due to slab pull, which does not exist in the Atlantic Plates.

**Acknowledgments**

We thank Nuno Bon de Sousa and Cleia Ribeiro for help in the preparation of this text.

**Figures**

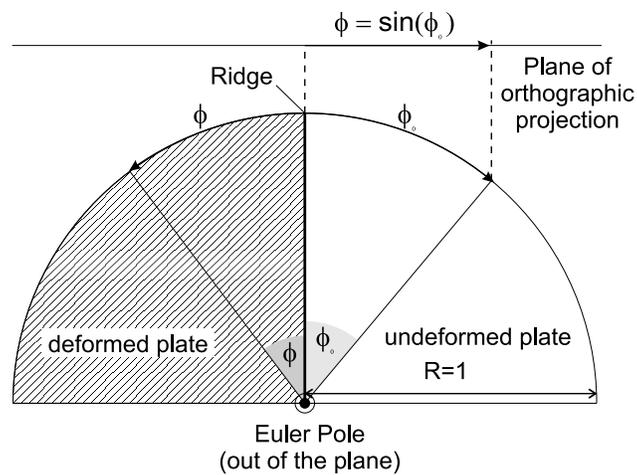

Figure 1 – Deformation of a spherical shell represented by the orthographic projection. Definition of the angles used in the text.

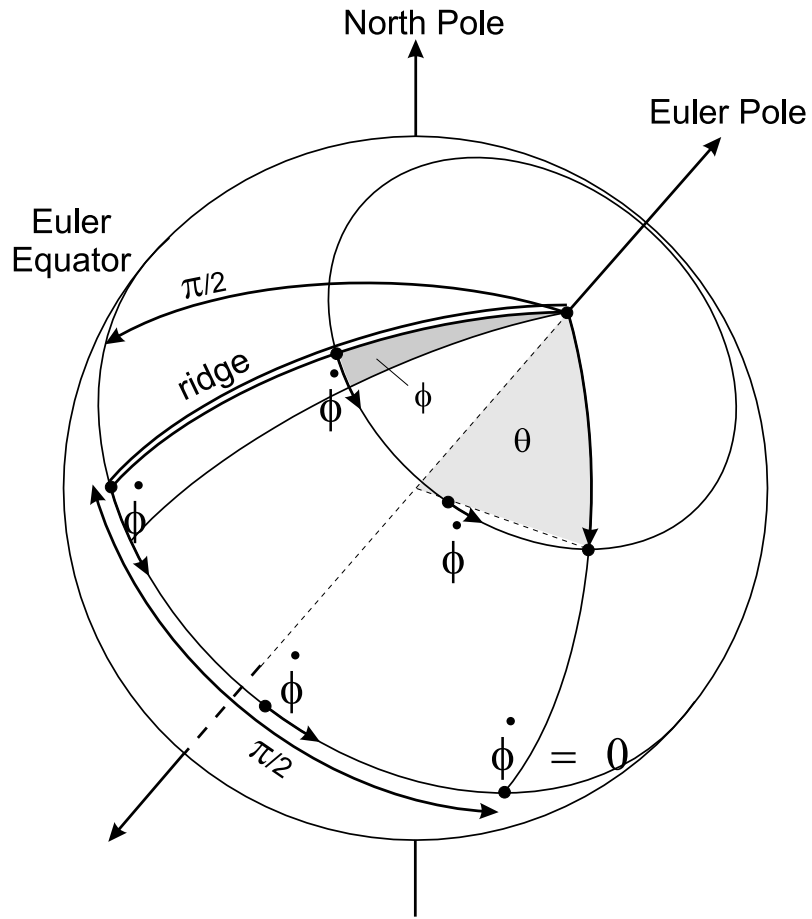

Figure 2 – The angular velocity vector $\dot{\phi}$ varies as a function of θ, the angular distance to the Euler Pole, and ϕ, the angular distance to the ridge. It becomes zero (a single dot) at the Euler Pole and at π/2 from the ridge.